# Laser scribing on HOPG for graphene stamp printing on silicon wafer

Jelena Butikova[*], Boris Polyakov, Lauris Dimitrocenko, Edgars Butanovs, Ivars Tale

Institute of Solid State Physics, University of Latvia, 8 Kengaraga St., Riga, LV-1063, Latvia

**Abstract.** Highly oriented pyrolytic graphite (HOPG) was scribed by pulsed laser beam to produce square patterns. Patterning of HOPG surface facilitates the detachment of graphene layers during contact printing. Direct HOPG-to-substrate and glue-assisted stamp printing of a few-layers graphene was compared. Printed graphene sheets were visualized by optical and scanning electron microscopy. The number of graphene layers was measured by atomic force microscopy. Glue-assisted stamp printing allows printing relatively large graphene sheets (40×40 µm) onto a silicon wafer, which can be important for microelectronics fabrication. The presented method is easier to implement and is more flexible than the majority of existing ways of placing graphene sheets onto a substrate.

## 1. Introduction

Graphene as a single monolayer of covalently bonded carbon atoms is an intriguing material for both fundamental and applied science. Due to its outstanding electronic and thermal properties (high charge carrier mobility, high heat conductance, etc.) graphene raised huge interest in the last decade, and became a real candidate as a successor of silicon in future microelectronics [1]. Yet, integration of graphene into electronic devices required its patterning and precise positioning.

There are plenty of graphene synthesis methods such as mechanical cleaving of highly pyrolytic graphite (HOPG) [2], epitaxial growth on silicon carbide [3], chemical vapour deposition of graphene on copper substrate [4, 5], chemical exfoliation and reduction of graphene oxide. The best electric properties, however, were demonstrated by graphene mechanically cleaved from HOPG. Cleaved graphene sheets, nevertheless, are usually irregular in shape and thickness, and therefore are not suitable for integration in microelectronics devices.

Graphene transfer stamp printing is becoming a popular method for controlled deposition of graphene sheets of predefined shape [6]. The main advantages of graphene transfer stamp printing technique are its flexibility, upwards scalability, and low cost. Moreover, this technique allows operating with the highest quality graphene derived from HOPG. There are two ways for preparing a graphene stamp. The first method of stamp preparation utilizes a silicon stamp with micromachined pillars (protrusions) having few tens of micrometers in diameter and several tens of micrometers in height, which is mechanically pressed onto HOPG, cutting graphene pieces adhered to pillars [7]. The disadvantages of this method are the need to prepare a stamp with micromachined pillars, which are rather brittle, and the expenses of the stamp production, which is quite high. An alternative method is direct laser or focused ion beam (FIB) scribing on HOPG surface. Still, utilization of FIB has a risk of graphene contamination by beam ions. Direct or mask-assisted laser scribing was successfully utilized for micropatterning of HOPG, multilayer graphene, and graphene oxide [8, 9]. In some investigations, however, non-flexible methods of micropatterning were used, and the following transfer of graphene elements to the substrate was of small effective area [10].

[*]Corresponding author: Institute of Solid State Physics, 8 Kengaraga St., Riga, LV-1063, Latvia.
Tel: +371 67187511, fax: +371 67132778. E-mail address: but@latnet.lv (J. Butikova)

In our experiments, squared configurations of graphene sheets were micropatterned by a nanosecond laser on HOPG for transfer stamp printing. Two different approaches of stamp printing were compared: direct stamping from patterned HOPG onto the silicon wafer and glue-assisted stamp printing.

**2. Materials and methods**

SPI-2 grade HOPG samples of were purchased from *SPI Supplies*. The second harmonic (532 nm) of Nd:YAG Q-switched nanosecond diode-pumped laser (NL200, *Ekspla*) with repetition rate of 500 Hz was used for HOPG scribing at ambient conditions. Laser beam was focused into 10-20 μm spot by an aspheric lens (*Thorlabs*) having a focal length of 11 mm. The sample was mounted on a motorized computer-controlled XY-stage (*Zaber Technologies*). A special LabView-based programme was written in order to control the parameters of the laser and the XY-stage at the same time.

Two methods of graphene printing were investigated. The schematics of these methods are shown in Figure 1. After the laser scribing, HOPG surface was wiped with a cotton swab or cleaved by the scotch tape to remove carbon debris deposited during the laser cutting of the grooves (Fig.1 c, d). In the first method, patterned HOPG was directly pressed with hands against a piece of a silicon wafer (*Semiconductor Wafer Inc.*) as it is shown in Fig.1 c. In the second method, a small drop of silicon universal glue (*Bison*) instead of widely used polydimethylsiloxane (PDMS) [11 - 13] was deposited onto the patterned HOPG surface and cleaved after 12 h (Fig.1 e). Graphite structures removed by the silicon glue were transferred onto the native oxidized silicon wafers (Fig.1 f).

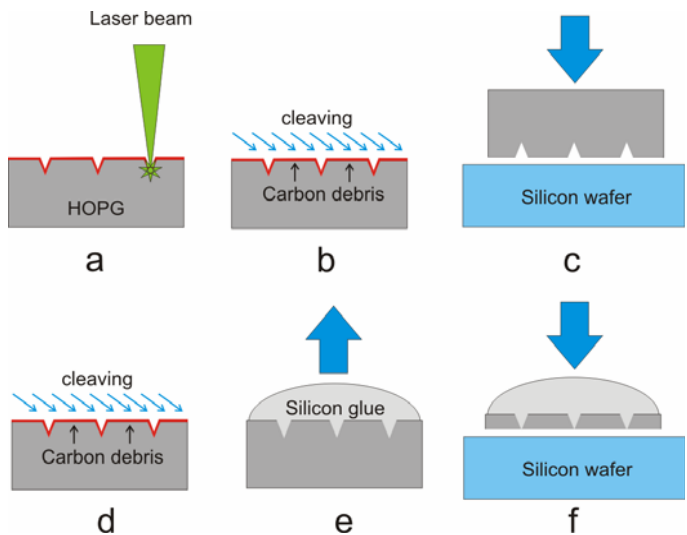

Figure 1. Schematics of two methods for graphene printing on the silicon wafer: direct (a-c) and glue-assisted (d-f) method. Laser scribing on HOPG surface (a). Cleaning of patterned HOPG surface (b, d). Direct transfer of graphene onto the silicon wafer (c). Graphene transfer onto the silicon wafer using silicon glue (e, f)

Optical microscope *Nikon* Eclipse L150 was used to investigate patterned HOPG surface and graphite structures stamped onto silicon wafer. Vega II (*Tescan*) scanning electron microscope (SEM) was used to investigate graphite structures transferred onto the silicon wafers. CP-II (*Veeco*) atomic force microscope (AFM) was used in tapping mode to measure the depth of the laser scribed grooves and the thickness of the transferred graphite structures.

## 3. Results and discussion

Squared micropatterns can be scribed by the focused nanosecond laser on the HOPG surface (Fig. 2). Squares were formed by scribing 2 perpendicular arrays of stripes with the distance of 40 microns between the stripes. Patterns were scribed by the laser beam with energy of 255 J cm$^{-2}$, and depth of the obtained grooves reached 1000 nm. With single laser shot, 25 nm of HOPG were cut. The surface of patterned HOPG samples was coated by the ablation products (amorphous and diamond-like carbon; see Fig. 2 c); thickness of the deposits was proportional to laser power. The surface containing this debris can be easily wiped with cotton swab or cleaved by an adhesive tape (Fig. 2 b, e). In general, cleaving by an adhesive tape results in a cleaner surface in comparison to wiping, however, cleaving introduces a risk of complete removal of the patterned areas. Cleaning procedure is absolutely necessary for the both printing methods. It is obviously necessary for the direct printing; otherwise, mostly the debris will be transferred to the substrate. In the case of the glue-assisted printing, the debris layer was much easier to detach from the silicon glue stamp, and single-layer graphene printing was not possible.

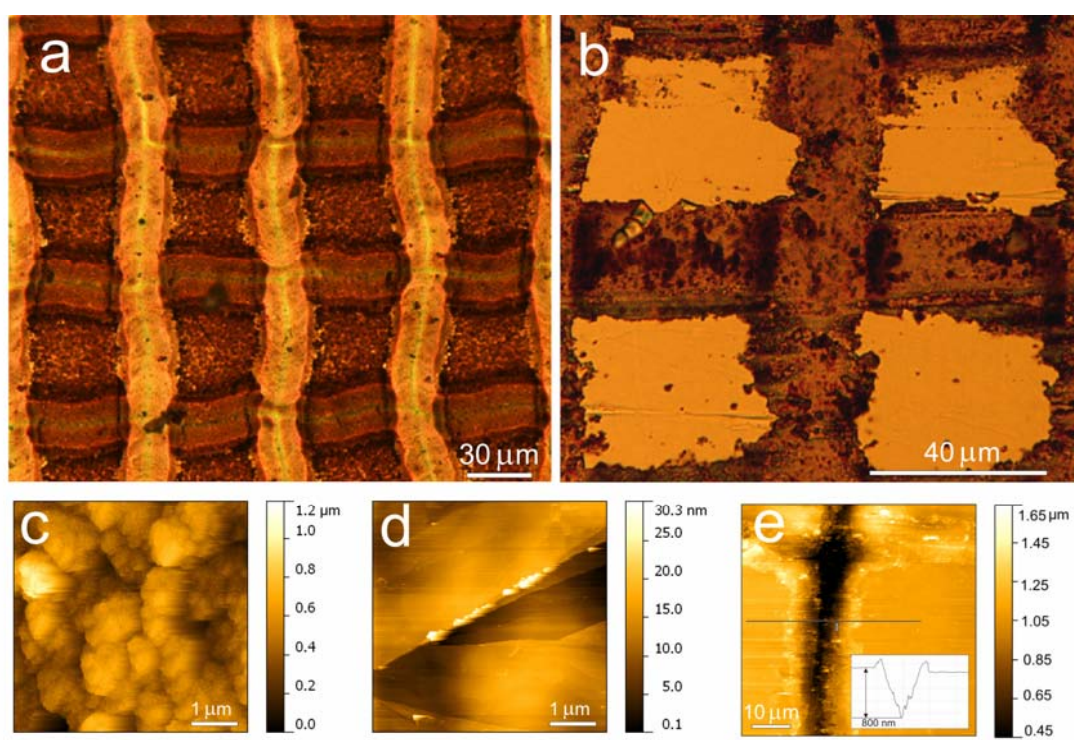

Figure 2. Optical microscope images of square patterns created by laser scribing on HOPG surface (a), and those after wiping with cotton swab (b). AFM image of the debris created during the process of laser scribing (c). AFM images of the surface covered by carbon debris (d) and cleaned surface (e)

Graphene can be printed onto an oxidized silicon wafer directly by pressing the wafer against the patterned HOPG sample. The productivity of this method was rather low. In the most cases, only very small graphite elements were transferred onto the wafer indicating borders of the scribed patterns, but without transfer of the large-area graphene sheets. The reason for this phenomenon to occur is the height – 100-200 nm – of the borders formed on the both sides if the grooves (see the cross-section measured in Fig. 2e). These borders are too high to be compressed during the process of direct printing, so they prevent graphene from transferring onto the surface of a substrate. In rare cases, however, graphene sheets were successfully printed on the silicon wafer. A few layers thick graphene sheets on oxidized silicon wafer can be visible in an optical microscope. The area of the transferred graphene sheets corresponds to the patterned feature size in general, but sometimes is smaller if crystallite border is smaller than the pattern size.

The visibility of graphene on the substrate in optical microscope depends on the number of graphene layers we wish to observe, and refractive index of the substrate material. Better contrast in optical microscope can be achieved placing graphene sheets on oxidized silicon wafer with specific silica thickness and applying optical filters [14]. In our experiments, thickness of the oxide layer was 125 nm, and the best contrast, and, hence, visibility was achieved with NCB 11 filter.

Even after applying a filter, graphene sheets thinner than 1-2 nm can still hardly be distinguished in optical microscope. In the best case, it is possible to recognize the outlines clearly, and only if the outlines are for a couple of nanometres higher than the monolayer graphene sheet. Conversely, scanning electron microscopy provides better contrast of single or a few-layer graphene printed on silicon wafers, as the whole area of printed graphene can be easily identified by SEM (Fig. 3a). In SEM images, darker regions correspond to thicker graphene layers. Multi-layer graphene appears as dark areas, while single-layer graphene appears like a very light shadow on wafer surface. One should be cautious during the electron microscopy of graphene, especially of single- or double-layer sheets, as the electron beam can not only alter the crystalline structure of graphene [15], but also introduce some damage.

It is also important to note that in SEM it is possible to achieve a good contrast and visibility of graphene if it is placed on an oxidized silicon wafers only. Graphene sheets transferred onto the pure silicon surface are almost invisible.

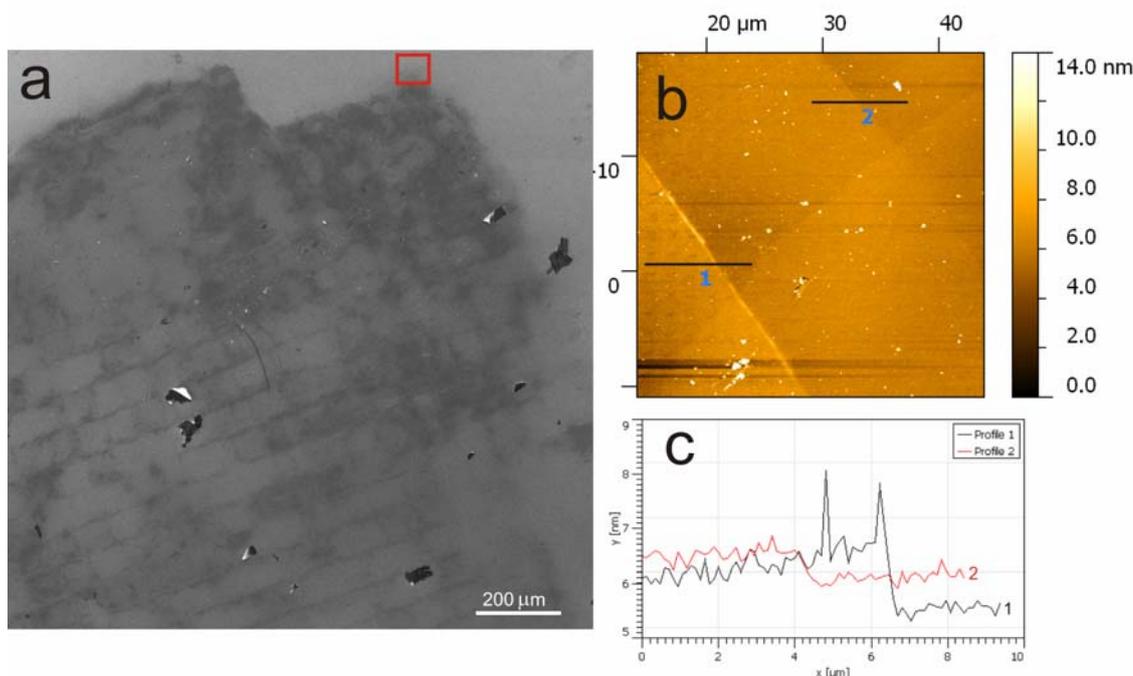

Figure 3. SEM image of stamp printed graphene sheets on oxidized silicon wafer (a). AFM image of the area marked by red rectangle on the SEM image (b). Cross sections demonstrating few- (Profile 1) and single-layer (Profile 2) graphene (c)

For quantification of printed graphene thickness produced by both methods, AFM was applied. The distance between the layers is equal to 0.34 nm, but the thickness of the single-layer graphene on silica is approximately 0.5 nm [16, 17]. Typical thickness of graphene printed by silicon glue-assisted stamp method was 0.5-2 nm, which corresponds to approx. 1-4 layers of graphene.

Silicon glue-assisted stamp printing produce better results in comparison to direct HOPG-to-silicon wafer printing in general. In the latter case, much higher pressure should be applied to bring HOPG and silicon wafer surface in contact due to low elasticity of the both materials. Silicon glue, in contrast, is very elastic, and adjusts itself easily to the plane of silicon wafer. We suppose that successive graphene transfer from stamp to substrate occurs only if some shear movement of stamp takes place relatively to the substrate surface. The layer of the glue should be thin enough, and the force applied to the stamp during the process of graphene transfer should not be excessive. Otherwise, graphene sheets might appear to be "smeared out" after the printing.

One more reason why silicon glue-assisted printing is more favourable method than HOPG-to-wafer direct printing is that the former method allows creating large-area graphene sheets. A sensitive aspect of silicon glue-assisted graphene printing is stamp lifting (Fig. 1e). Sometimes, during the lifting of the layers cut by the laser, some non-patterned graphene layers happen to be lifted by the stamp. This is highly unwanted phenomena caused by macroscopic defects in HOPG. However, upper non-patterned layer can be cleaved from the stamp surface by an adhesive tape.

## 4. Conclusions

Squared configurations of graphene sheets were micropatterned by a nanosecond laser on HOPG for transfer stamp printing. Two approaches of stamp printing were investigated: direct stamping from patterned HOPG onto the silicon wafer and glue-assisted stamp printing. The second approach was proven to be valid for printing a single- or few-layer graphene sheets (with corresponding thickness of 0.5-2 nm) having an area of 40×40 microns on silicon wafer substrate. Silicon glue was used for printing instead of PDMS.

The key to the successive transfer of graphene sheets onto the silicon substrate seems to be shear movement occurring during the pressing a drop of a glue to the micropatterned HOPG surface. The thickness of the glue and the pressing force at the moment of printing are important parameters as well.

The main advantages of the proposed glue-assisted stamp printing for the transfer of graphene sheets are its flexibility and relatively easy implementation, and, as a result, the possibility of transferring of at least 4 times larger graphene sheet – 40 microns compared to 10 microns reported in, for instance, [7-9] – keeping the same graphene thickness from single to few layers.

**Acknowledgements**

This investigation was supported by the National Research Programme IMIS, section "Graphene".

The authors are grateful to Andris Voitkans for the development of a programme for laser control and XY-stage movement.